\documentclass[twocolumn]{aastex631}

\usepackage{amssymb, amsfonts, amsmath,framed}
\usepackage{float,bm}
\usepackage{soul}

\begin{document}

\title{Intermittency and Dissipative Structures Arising from Relativistic Magnetized Turbulence}


\correspondingauthor{Zachary Davis}
\email{zkd@purdue.edu}

\author[0000-0002-0959-9991]{Zachary Davis}
\affiliation{Department of Physics, Purdue University, 525 Northwestern Avenue, West Lafayette, IN 47907, USA}

\author[0000-0001-8822-8031]{Luca Comisso}
\affiliation{Department of Astronomy and Columbia Astrophysics Laboratory, Columbia University, New York, NY 10027, USA}

\author[0000-0003-1503-2446]{Dimitrios Giannios}
\affiliation{Department of Physics, Purdue University, 525 Northwestern Avenue, West Lafayette, IN 47907, USA}

\begin{abstract}

Kinetic simulations of relativistic turbulence have significantly advanced our understanding of turbulent particle acceleration. Recent progress has highlighted the need for an updated acceleration theory that can account for particle acceleration within the plasma's coherent structures. Here, we investigate how intermittency modeling connects statistical fluctuations in turbulence to regions of high energy dissipation. This connection is established by employing a generalized She-Leveque model to characterize the exponents $\zeta_p$ for the structure functions $S^p \propto l^{\zeta_p}$. The fitting of the scaling exponents  provide us with a measure of the co-dimension of the dissipative structures, for which we subsequently determine the filling fraction. We perform our analysis for a range of magnetizations $\sigma$ and relative fluctuation amplitudes ${\delta B_0}/{B_0}$. We find that increasing values of $\sigma$ and ${\delta B_0}/{B_0}$ allow the turbulent cascade to break sheet-like structures into smaller regions of dissipation that resemble chains of flux-ropes. However, as their dissipation measure increases, the dissipative regions become less volume filling. With this work we aim to inform future turbulent acceleration theories that incorporate particle energization from interactions with coherent structures within relativistic turbulence. 






\end{abstract}


\section{Introduction} \label{sec:intro}

A central goal in high-energy astrophysics is to uncover the physical processes that power the most extreme particle accelerators, which are accountable for bright electromagnetic observations. Often, as is the case with gamma-ray bursts (GRBs) and jets from active galactic nuclei (AGNs), the sources are observed to have a broad electromagnetic spectrum, whose interpretation requires a relativistic non-thermal distribution of particles undergoing radiative cooling \citep{ghisellini1998,band1993}. The mechanisms responsible for the rapid particle acceleration needed to produce these non-thermal particle distributions are a subject of ongoing debate (see \citet{mathews2020} for a recent review). Nonetheless, it is commonly assumed that the free energy required for particle acceleration comes from large-scale perturbations within the jet, that can lead to the formation of shocks \citep[e.g.,][]{bottcher2010} or large-scale magnetic reconnection layers \citep[e.g.,][]{giannios2013}. Regardless of the specific scenario, a large scale separation exists from the energy injection scale to plasma kinetic scales, where energy dissipation occurs. This scale separation typically involves turbulence modulation, highlighting the need to understand turbulence for a comprehensive understanding of particle acceleration and high-energy emission.

Turbulence, a complex nonlinear phenomenon already challenging the study in hydrodynamical fluids on Earth, poses additional challenges in astrophysical plasmas. Factors such as low particle collisionality, strong magnetic fields, relativistic velocities, and extreme temperatures contribute to the complexity. The analytical treatment of turbulence often relies on phenomenological cascade models (or scaling theories) that link large-scale fluid fluctuations to smaller scales, extending down to the dissipation scale.
The most renowned model of hydrodynamic turbulence is due to  \citet{kolmogorov1941}, with adaptations later developed for magnetohydrodynamics (MHD) \citep[e.g.,][]{iroshnikov1963,kraichnan1965,goldreich1995}. Initial investigations into the validity of cascade models were conducted through numerical simulations using the MHD approximation. However, recent progress in this field has seen the development of first-principle kinetic simulations, which are crucial for enhancing our understanding of turbulence in weakly collisional plasmas, particularly at the smallest scales where most energy is dissipated.

Given its relevance in relativistic magnetized outflows, we focus on investigating turbulence in conditions where the magnetic energy density exceeds the rest-mass energy density and the plasma pressure. Under these conditions, plasma motions approach the speed of light, marking the \textit{relativistic turbulence} regime. This regime of turbulence is intrinsically linked to magnetic reconnection \citep{lazarian2012,comisso2019}, which has been shown to be an efficient accelerator of relativistic particles \citep[e.g.,][]{comisso2018}. Additionally, in the relativistic regime, the stochastic Fermi acceleration mechanism \citep{fermi1949} operates on short timescales, facilitating the rapid conversion of magnetic energy into plasma particle energy. Turbulent acceleration in magnetized plasmas has been associated with various sources of high-energy emissions, including solar flares \citep{miller1996}, gamma-ray bursts \citep{bykov1996}, and blazar jets \citep{marscher2014,davis2022,Zhang2023}.

The traditional analytical approach to understanding turbulence's role in particle acceleration is through quasi-linear theory (QLT) \citep{bernstein1966}. In QLT, particle trajectories are gyro-averaged before scattering off plasma waves, and the analytical description often involves calculating a diffusion tensor derived from linear eigenmodes of the plasma's MHD waves \citep[see, e.g.,][]{demidem2020}. However, large-amplitude turbulent fluctuations invalidate QLT, and recent fully kinetic particle-in-cell (PIC) simulations have indeed cast doubts on the QLT description. In the relativistic turbulence regime, QLT is unable to account for the observed development of pitch-angle anisotropy \citep{comisso2020,Comisso21PRL}, recreate the observed particle distributions without unexplained advection coefficients \citep{zhdankin2020}, or provide a mechanism for the observed acceleration at current sheets \citep{comisso2019}. QLT's inability to describe recent results from PIC simulations has spurred the search for new analytical descriptions of particle acceleration in relativistic turbulence. \citet{lemoine2021} suggested abandoning the idea that particle acceleration in relativistic turbulence is due to wave interactions and rather that acceleration may be mainly do to a collection of interactions with discrete structures within turbulence. Current sheets are of particular interest because of their common occurrence in magnetized turbulence \citep{comisso2019,zhdankin2020} and, when these sheets undergo reconnection, they efficiently dissipate magnetic energy. Current sheets also provide a link to phenomenological cascade models incorporating dissipative structures \citep{she1994,dubrelle1994,biskamp2000} that can be associated with the observed intermittency, pitch-angle anisotropy and energy spectra \citep{comisso2019,zhdankin2020}.

To bridge the gap between phenomenological theories of intermittency and the dissipative structures responsible for particle energization, we analyze PIC simulations of relativistic magnetized turbulence. Our investigation begins with the characterization of the turbulence intermittency and its dependence on two key physical parameters: the plasma magnetization and the amplitude of the magnetic fluctuations with respect to the mean magnetic field. We employ a general log-Poisson model of turbulence to establish a link between the coherent structures and the phenomenological model. Once established, we investigate the filling fractions of these coherent structures. This approach is repeated for different values of plasma magnetization and level of the magnetic fluctuations to analyze the dependence of the dissipative structure properties on these key plasma parameters. Finally, we discuss the implications for current particle acceleration theories before concluding.

\section{Numerical Setup} \label{sec:sim_setup}

To investigate the statistics of relativistic magnetized turbulence from a first-principles standpoint, we solve the Vlasov–Maxwell system of equations through the Particle-in-Cell (PIC) method \citep{birdsall_langdon_85} using the publicly available code TRISTAN-MP \citep{buneman_93, spitkovsky_05}. We perform the numerical simulations in a triply periodic cubic domain $L^3$ that is discretized into a regular lattice of $1024^3$ cells. 
We initialize a uniform electron-positron plasma with a total particle density of $n_0$ according to a Maxwell-J\"{u}ttner distribution with dimensionless temperature $\theta_0 = {k_B T_0}/{m c^2} = 0.3$. Here, $T_0$ is the initial plasma temperature,  $k_B$ indicates the Boltzmann constant, $m$ is the electron mass, and $c$ is the speed of light in vacuum. Turbulence is seeded by initializing a spectrum of magnetic fluctuations having polarizations transverse to a uniform mean magnetic field  $\langle{\bm{B}}\rangle=B_0{\bm{\hat z}}$ (see \citet{comisso2018,comisso2019} for details). The initial magnetic energy spectrum peaks near $k_p = 6 \pi /L$, which defines the energy-carrying scale $l_0 = 2 \pi /k_p$ and limits the particles maximum lorentz factor through stochastic acceleration \citep{comisso2018}.

The strength of initial fluctuating magnetic energy relative to plasma enthalpy is quantified by 
\begin{equation}
\label{eq:initial_magnetization}
\sigma =\frac{\delta B_0^2}{4\pi n_0 w_0 m c^2} \, ,
\end{equation}
where $\delta B_0 = \langle {\delta {B^2} (t=0)} \rangle^{1/2}$ is the initial rms amplitude of the magnetic field fluctuations and $w_0 =\left[K_3\left(1 / \theta_0\right) / K_2\left(1 / \theta_0\right)\right]$ is the initial enthalpy per particle, with $K_n(z)$ indicating the modified Bessel function of the second kind of order $n$. 
The corresponding total magnetization is $\sigma + \sigma_{B_0} = (\delta B_0^2 + B_0^2)/{4\pi n_0 w_0 m c^2}$. Since our work focuses on studying \textit{relativistic} plasma turbulence, characterized by Alfv{\'e}nic velocity fluctuations ${v_{A}} = c\sqrt {{\sigma}/(1 + {\sigma})}  \sim c$, we focus on the magnetization regime $\sigma \gg 1$. 
Our simulations cover a range of magnetizations with values $\sigma \in \left\{ {2.5,5,10,20,40} \right\}$, 
and different strengths of the mean magnetic field, corresponding to ratios $\delta B_{0}/B_0 \in \left\{ {0.5,1,2} \right\}$.

We adopt a spatial resolution of $\Delta x = d_{e0}/3$, implying $L = 1024d_{e0}/3$, where $d_{e0}=c/\omega_{p0}$ indicates the initial plasma skin depth and $\omega_{p0} =\sqrt {4\pi n_0 {e^2}/\gamma_{th0} {m}}$ is the relativistic plasma frequency, where $ \gamma_{t h 0} = w_0 - \theta_0$ is the initial mean thermal Lorentz factor. We employ an average of 4 computational particles per cell. Earlier studies \citep{comisso2018,comisso2019} have demonstrated convergence with respect to these numerical parameters. Since we are interested in studying fully-developed turbulence, we ran so the simulation up to $t \sim 3 l_0/c$, at which point, the turbulence has been fully developed. In the following, we mainly discuss results obtained from the magnetic field $b = B/B_0$, the current density $j = {J}/e n_0 c$, and the fluid bulk velocity $v = V/c$, obtained by averaging the velocities of individual particles.

\section{Intermittency Model} 

Turbulence is characterized by a broad range of scales linking energy initially injected at large scale $l_0$ down to a much smaller scale $l_d \ll l_0$ where energy can be efficiently dissipated. To this day, there is no fully developed theory that comprehensively explains this process. Rather, the best results in explaining the energy cascade have come from building a collection of hypotheses that are then integrated into a phenomenological model.

One of the most celebrated models is Kolmogorov's (referred to as K41) \citep{kolmogorov1941}, which established the phenomenological energy spectrum $E(k) \propto k^{-5/3}$ and introduced the pivotal ``4/5 law''. This law gives the third moment of velocity fluctuations as $\langle\left(\delta v(l)\right)^3\rangle=-\frac{4}{5} \epsilon l$, where $\epsilon$ is the cascading energy flux, $l$ is a given eddy size and
\begin{equation}
\label{eq:SF_def}
    \delta v(l) \equiv v( \bm{r}+ \bm{l}) - v( \bm{r}) 
\end{equation}
is the velocity fluctuation. Kolmogorov suggested introducing a simple self-similarity assumption to extend this law to a higher power of $p$. In this case, one has 
\begin{equation}
    S^{p} (l) = \left\langle\left(\delta v(l)\right)^p\right\rangle  \propto l^{\zeta_p} 
\end{equation}
with $\zeta^{K41}_p = p/3$ for K41. Though initially successful for hydrodynamical turbulence, deviations from the scaling relation become evident for $p>3$.

General deviations in $\zeta_p$ from $p/3$ are often attributed to intermittency, or the tendency for subsequent eddys to become less volume filling than the previous generation. \citet{she1994} proposed a phenomenological model to account for the observed deviations in hydrodynamical turbulence by introducing a recursion relation that resulted in a hierarchy of dissipative structures. From this hierarchy, \citet{she1994} derived the relation
\begin{equation}
    \label{eq:zeta_SL}
    \zeta_p=(p / g)(1-x)+C_0\left[1-\left(1-x / C_0\right)^{p / g}\right] 
\end{equation}
for the scaling exponents. Here, Eq. (\ref{eq:zeta_SL}) is presented as it is in \citet{politano1995}, where the physical assumptions resulting in no free parameters have been removed. 
In Eq. (\ref{eq:zeta_SL}), $g$ corresponds to fluctuation scaling (i.e, for velocity, $\delta v_l \propto l^{1/g}$),  $x$ relates to cascade time scale of the dissipative structures, $\tau_{l} \propto l^x$,  and $C_0$ is the co-dimension of the structures, related to the fractal dimension by $C_0 = d  - D$. Here $d$ is the embedded space dimension (i.e, $d=2$ for 2D turbulence and $d=3$ for 3D turbulence) and $D$ is the fractal dimension of the structures.

In \citet{dubrelle1994}, it was shown that the model presented in \citet{she1994} corresponds to the dissipative structures processing a distribution described by a log-Poisson equation and in the correct limits, can be described with a random fractal $\beta$ model.  In the general form presented in Eq. (\ref{eq:zeta_SL}), we are able to compare different phenomenological models of turbulence. A K41 scaling can be recovered with $x=0$, $g=3$ and $C_0 =0 $.  To reproduce the exact result in \citet{she1994}, let $x=2/3$, $g=3$, and since \citet{she1994} considered filamentary structures, $C_0=2$.

Including magnetic fields also has expected effects on the parameters in Eq. (\ref{eq:zeta_SL}). Following \citet{iroshnikov1963,kraichnan1965}, we can expect $g=4$ due larger cascade time from the Alfvén wave interactions. \citet{goldreich1995} introduced anisotropy of the magnetic field fluctuations into the picture. This finds $g=3$ for fluctuations perpendicular to the background magnetic field and $C_0 = 1$ by assuming that dissipation occurs in sheets. \citet{MB2000PRL} proposed to modify the \citet{she1994} model by simply setting $C_0 = 1$ for magnetized turbulence, as turbulent MHD simulations were abundant in electric current sheets.

Instead of presuming the structure of the turbulent fluctuations, in the following we analyze the simulations discussed in Section \ref{sec:sim_setup} using structure function statistics as a tool to characterize intermittency, examine the applicability of previous phenomenological models, and understand how these models relate to the most dissipative structures in turbulence.

\section{Structure's Co-Dimensionality}  \label{sec:struct_func}

\begin{figure}
    \centering
    \includegraphics[width=0.45\textwidth]{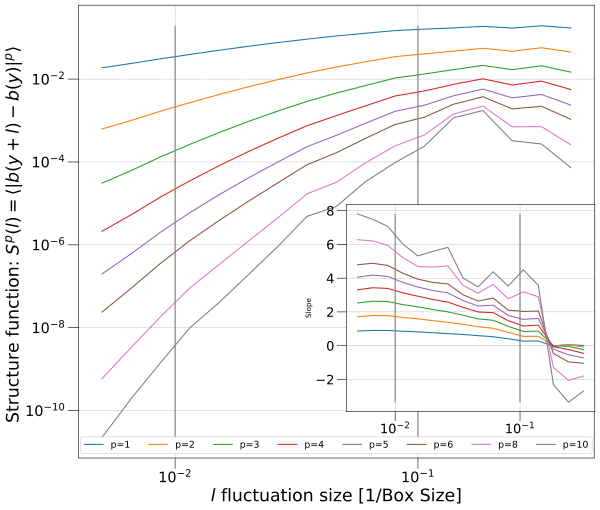}
    \caption{SF of fluctuations in $b$ for $p$ in the range 1-10. Colors denote different values of $p$. The black vertical lines denote the scales $L = 1/100$ and $L=1/10$ as fractions of the simulation box size, where the SF is fitted to the power law $\propto l^{\zeta_p}$. The inset shows the slope of the SF.}
    \label{structure_functions_sf}
\end{figure}

In order to carry out an analysis that can then be compared to phenomenological models, we first outline the construction of the structure function (SF) . For simplicity, the analysis is carried out using the magnitude of fluctuation, as seen for the fluid velocity in Eq. \ref{eq:SF_def}. Due to observed anisotropy in the turbulent spectrum\citep{zrake2013,muller2003}, we also analyze the parallel component of a fluctuation by letting $ \bm{l}$ go to $\bm{l_{\parallel}}$, referring to fluctuation component parallel to the mean magnetic field, namely $ \bm{B_0}$
\begin{equation}
    \delta v(l)_\parallel \equiv \left[ \bm{v}( \bm{r}+ \bm{l})- \bm{v}( \bm{r})\right] \cdot \frac{ \bm{l}}{l} \, .
\end{equation}
To calculate the value of a given fluctuation for increment $l$, the simulation grid is split up into 2D slices  of thickness $l$  along the given $ \bm{r}$  value. If a slice does not fall on a grid point, the values are linearly interpolated. The slice is then subtracted from another slice separated by the distance $l$. We repeat this for a 150 unique, randomly selected pairs of slices. In order to build a SF with this fluctuation, all subtracted pairs have the absolute value taken, raised to the $p$ before averaging. An example  of the produced SF is shown in Figure \ref{structure_functions_sf}.

To compare and fit to phenomenological models, we calculate the SF exponent $\zeta_p$ by fitting the SF to a power law $l^{\zeta_p}$. The fitting range is set from $1/100$ to $1/10$ of the box size to ensure we are consistently within the inertial range. This  range is shown in Figure \ref{structure_functions_sf}. We use a least squares method for the fit, and the coefficient's error is determined from the covariance matrix.  We calculate $\zeta_p$ values for orders $p = 1, ..., 10$ and for quantities $\delta b$, $\delta j$, $\delta v$.  All measurements are then repeated for each simulation.  Additionally, we fit Eq. (\ref{eq:zeta_SL}) for $C_0$ using again a least squares fit. Parameters $x=2/3$ and $g=3$ are assumed constant. These results are detailed in Table 1.

\begin{table}[]
\label{tab:C0_results}
\begin{tabular}{lcccc}
                         & $C_0$           & $C_{0,\parallel}$ & $\sigma$ & ${\delta B_0}/{B_0}$ \\ \hline
\multicolumn{1}{l|}{$\delta b$} & $0.963\pm0.06$  & $1.14\pm0.08$       & 2.5      & 1                      \\
\multicolumn{1}{l|}{$\delta b$} & $1.0\pm0.08$    & $1.3\pm0.08$        & 5        & 1                      \\
\multicolumn{1}{l|}{$\delta b$} & $1.09\pm0.04$   & $1.11\pm0.03$       & 10       & 0.5                    \\
\multicolumn{1}{l|}{$\delta b$} & $1.21\pm0.06$   & $1.76\pm0.05$       & 10       & 1                      \\
\multicolumn{1}{l|}{$\delta b$} & $1.23\pm0.08$   & $2.98\pm0.02$       & 10       & 2                      \\
\multicolumn{1}{l|}{$\delta b$} & $1.51\pm0.04$   & $1.94\pm0.09$       & 20       & 1                      \\
\multicolumn{1}{l|}{$\delta b$} & $1.58\pm0.09$   & $2.2\pm0.1$         & 40       & 1                      \\
\multicolumn{1}{l|}{$\delta j$} & $0.7\pm0.06$    & $0.683\pm0.06$      & 2.5      & 1                      \\
\multicolumn{1}{l|}{$\delta j$} & $0.753\pm0.05$  & $0.685\pm0.06$      & 5        & 1                      \\
\multicolumn{1}{l|}{$\delta j$} & $0.718\pm0.01$  & $0.68\pm0.03$       & 10       & 0.5                    \\
\multicolumn{1}{l|}{$\delta j$} & $0.805\pm0.05$  & $0.701\pm0.06$      & 10       & 1                      \\
\multicolumn{1}{l|}{$\delta j$} & $0.796\pm0.03$  & $0.694\pm0.03$      & 10       & 2                      \\
\multicolumn{1}{l|}{$\delta j$} & $0.824\pm0.08$  & $0.711\pm0.06$      & 20       & 1                      \\
\multicolumn{1}{l|}{$\delta j$} & $0.954\pm0.05$  & $0.728\pm0.07$      & 40       & 1                      \\
\multicolumn{1}{l|}{$\delta v$} & $1.08\pm0.02$   & $2.28\pm0.01$       & 2.5      & 1                      \\
\multicolumn{1}{l|}{$\delta v$} & $1.187\pm0.008$ & $1.98\pm0.06$       & 5        & 1                      \\
\multicolumn{1}{l|}{$\delta v$} & $1.03\pm0.02$   & $1.59\pm0.02$       & 10       & 0.5                    \\
\multicolumn{1}{l|}{$\delta v$} & $1.16\pm0.01$   & $2.05\pm0.05$       & 10       & 1                      \\
\multicolumn{1}{l|}{$\delta v$} & $0.817\pm0.02$  & $2.19\pm0.04$       & 10       & 2                      \\
\multicolumn{1}{l|}{$\delta v$} & $0.997\pm0.02$  & $1.59\pm0.06$       & 20       & 1                      \\
\multicolumn{1}{l|}{$\delta v$} & $0.865\pm0.03$  & $1.69\pm0.04$       & 40       & 1                     
\end{tabular}
\caption{A summary table of the intermittency results. For each variable ($\delta b$, $\delta j$, $\delta v$), the co-dimensions $C_0$ and $C_{0,\parallel}$ are found by fitting Eq. (\ref{eq:zeta_SL}) for values of $\sigma=2.5,5,10,20,40$ and ${\delta B_0}/{B_0}=0.5,1,2$.}
\end{table}

\begin{figure}
  \centering
  \begin{tabular}{@{}c@{}}
    \includegraphics[width=\linewidth]{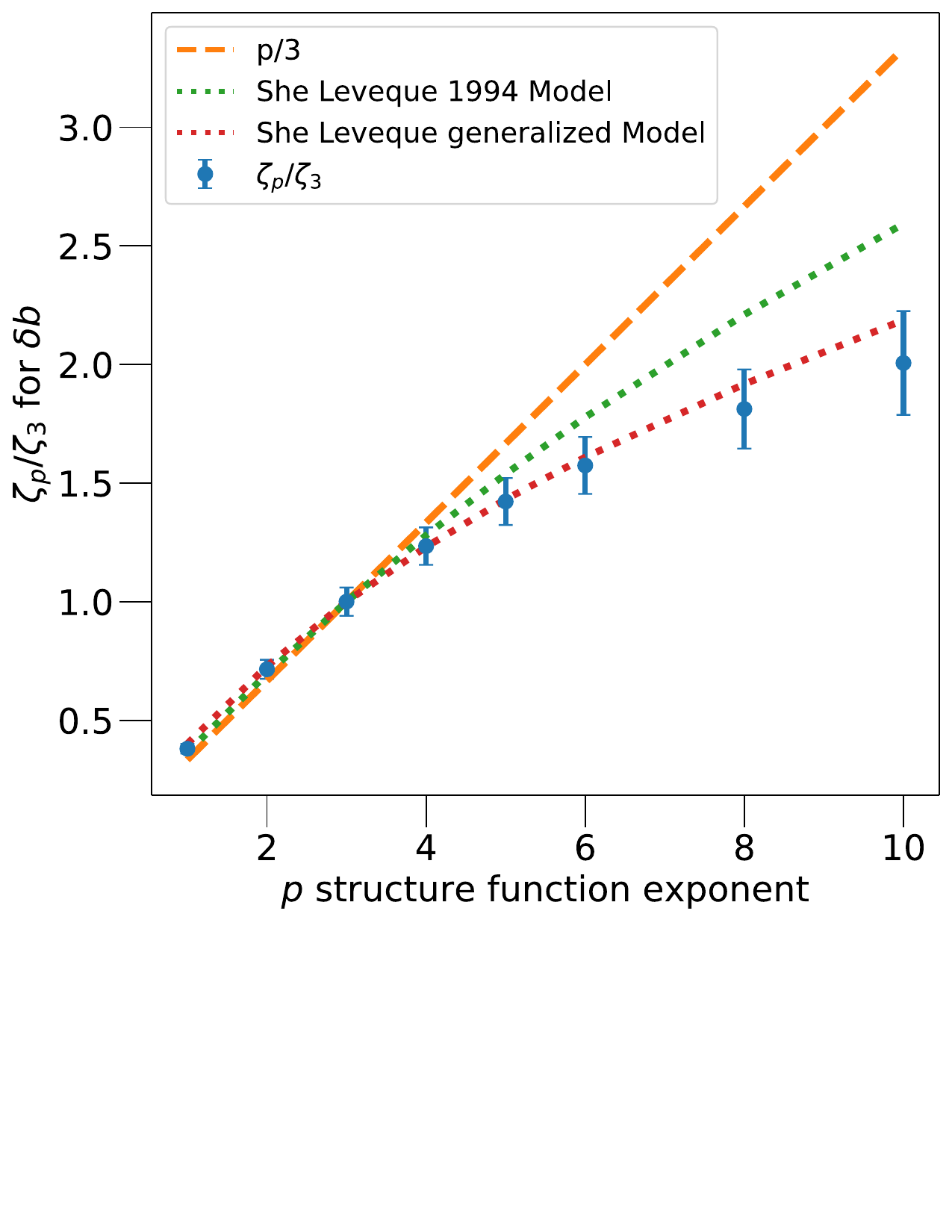} \\[\abovecaptionskip]
  \end{tabular}

  \vspace{-3.5cm}

  \begin{tabular}{@{}c@{}}
    \includegraphics[width=\linewidth]{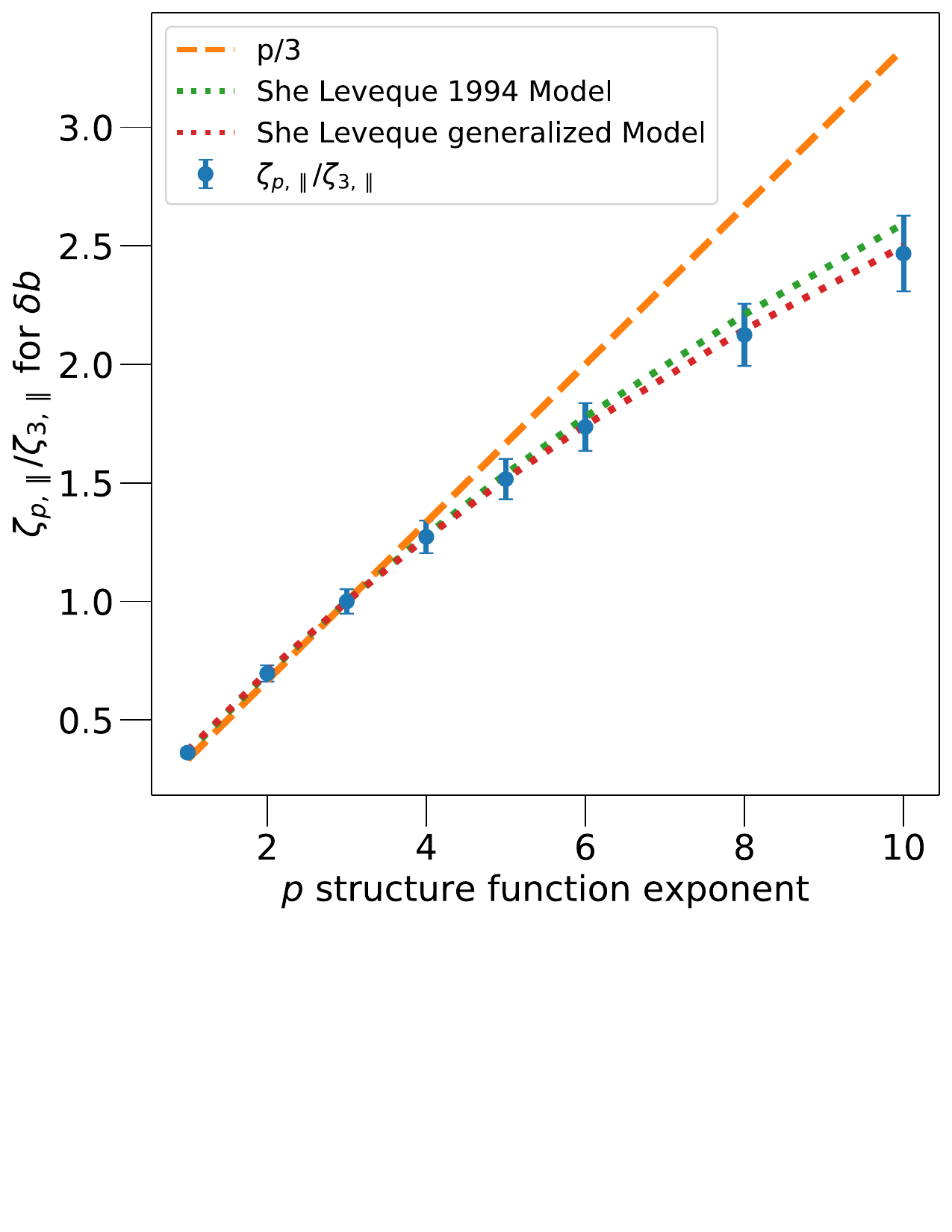} \\[\abovecaptionskip]
  \end{tabular}
    \vspace{-3cm}
  \caption{SF exponents $\zeta_p$ vs $p$ for $\sigma=10$ and ${\delta B_0}/{B_0}=1$. Compared with the trends from \citet{kolmogorov1941}(orange dashed line) and the result from \citet{she1994}(green dotted line). The red dotted line shows our fit to data using Eq. (\ref{eq:zeta_SL}) with $x=2/3$ and $g=3$. $C_0=1.14$ for the top figure and $C_0=1.73$ for the bottom figure.  }
  \label{fig:cpf_example_fig}
\end{figure}

For illustrative purposes, we show in Figure \ref{fig:cpf_example_fig} the general appearance of the scaling exponents for the magnetic field both from the perpendicular and parallel SF.
The pronounced intermittency is evident through the large deviation from the K41 scaling after $p=3$. The red dotted line shows the results from \citet{she1994}, and we can see that we generally have a significant deviation demanding lower values for the co-dimension $C_0$ than those typically found in hydrodynamic turbulence. The parallel direction displays smaller deviations that are closer to the original \citet{she1994} value. The smaller deviation in the parallel direction is consistent with what is seen in \citet{zrake2013}.

\begin{figure}
  \centering
  \begin{tabular}{@{}c@{}}
    \includegraphics[width=\linewidth]{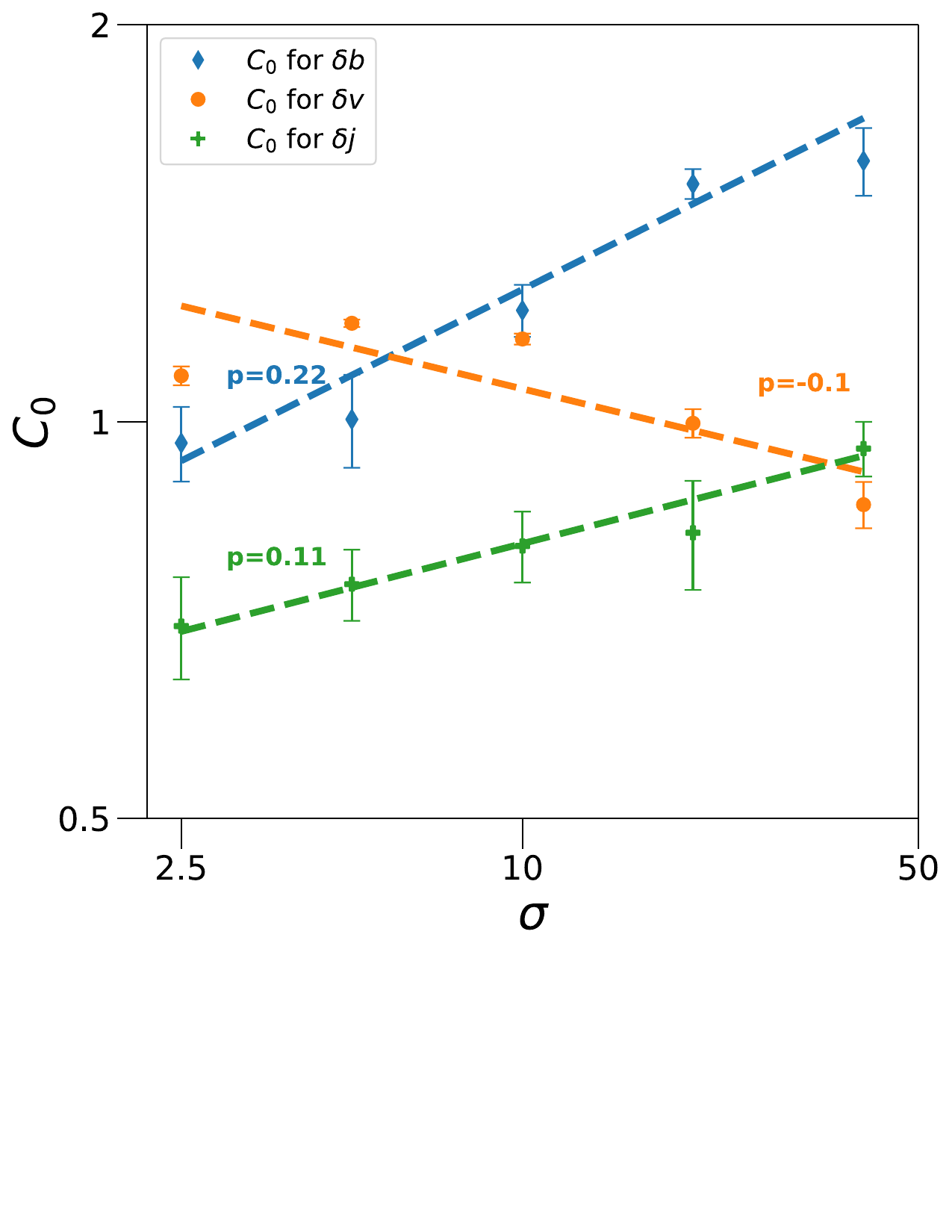} \\[\abovecaptionskip]
  \end{tabular}

  \vspace{-3.0cm}
  
  \begin{tabular}{@{}c@{}}
    \includegraphics[width=1.0\linewidth]{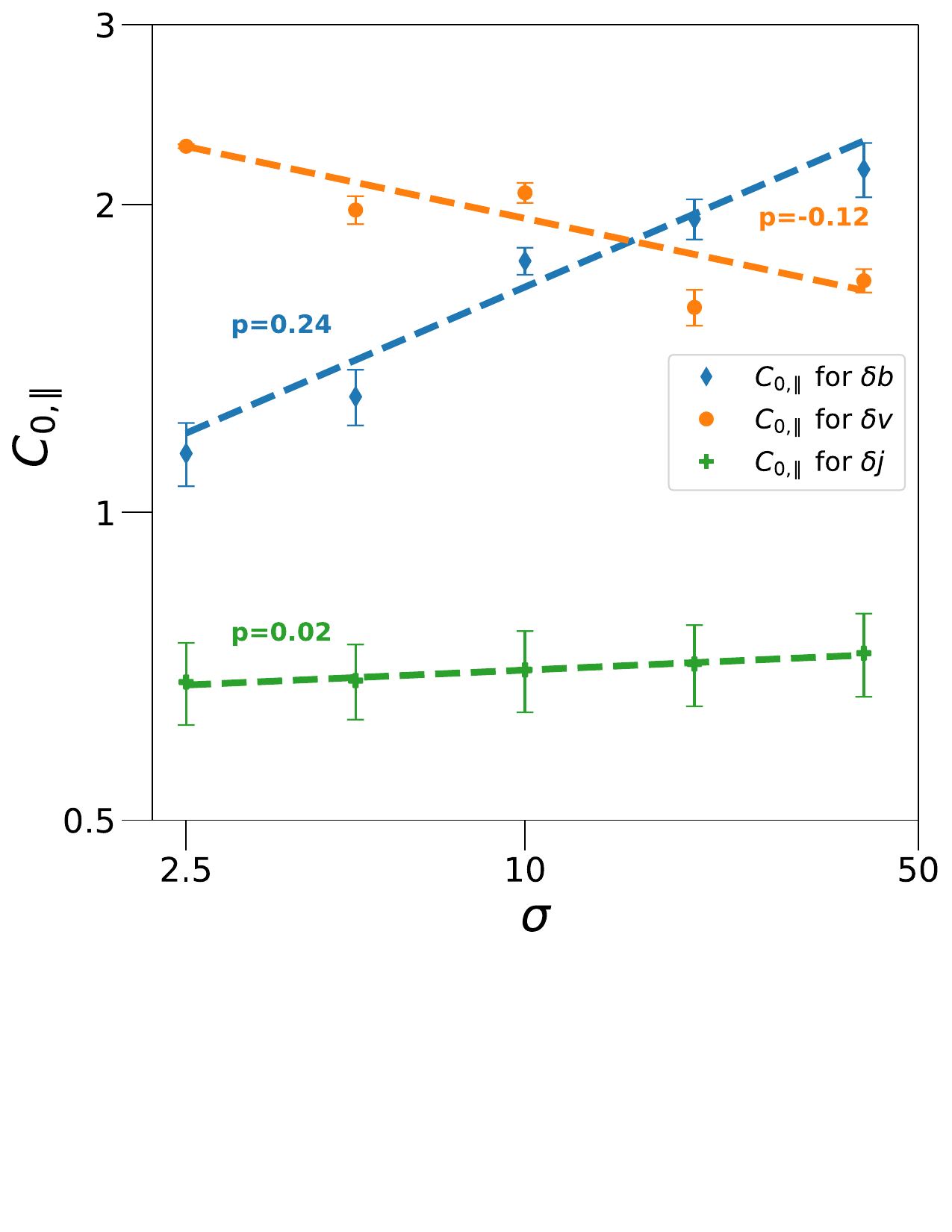} \\[\abovecaptionskip]
    
  \end{tabular}
\vspace{-3.0cm}
  \caption{Dependence of the co-dimensions $C_0$ and $C_{0,\parallel}$ on $\sigma$. Orange, blue, and green denote fluctuations in $\delta v$, $\delta b$, and $\delta j$, respectively. Dashed lines show a power law with a similar slope to data of the corresponding color. Top and bottom figures are in log-log space. }
  \label{fig:c0_vs_sigma}
\end{figure}

\begin{figure}
  \centering
  \begin{tabular}{@{}c@{}}
    \includegraphics[width=\linewidth]{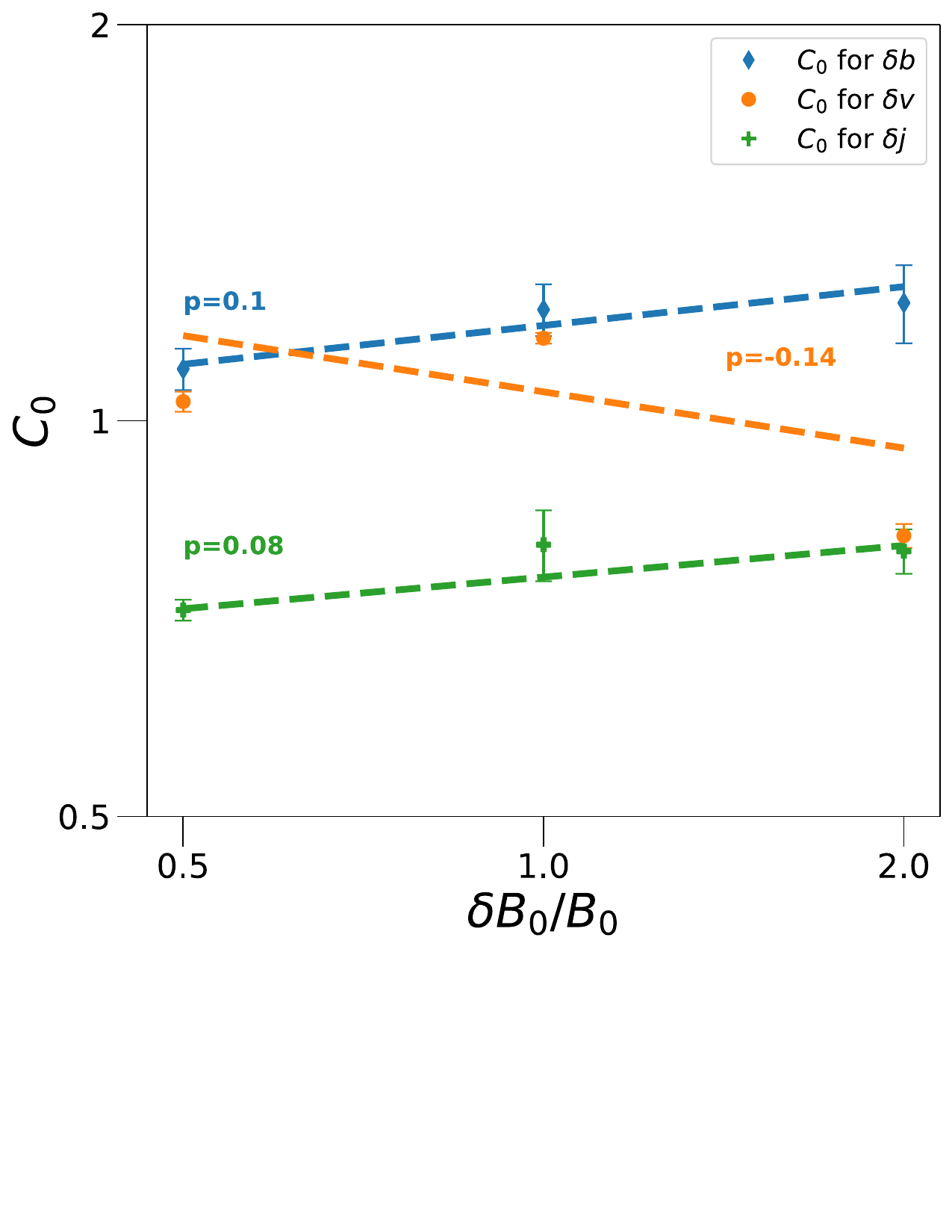} \\[\abovecaptionskip]
  \end{tabular}

  \vspace{-3.0cm}

  \begin{tabular}{@{}c@{}}
    \includegraphics[width=\linewidth]{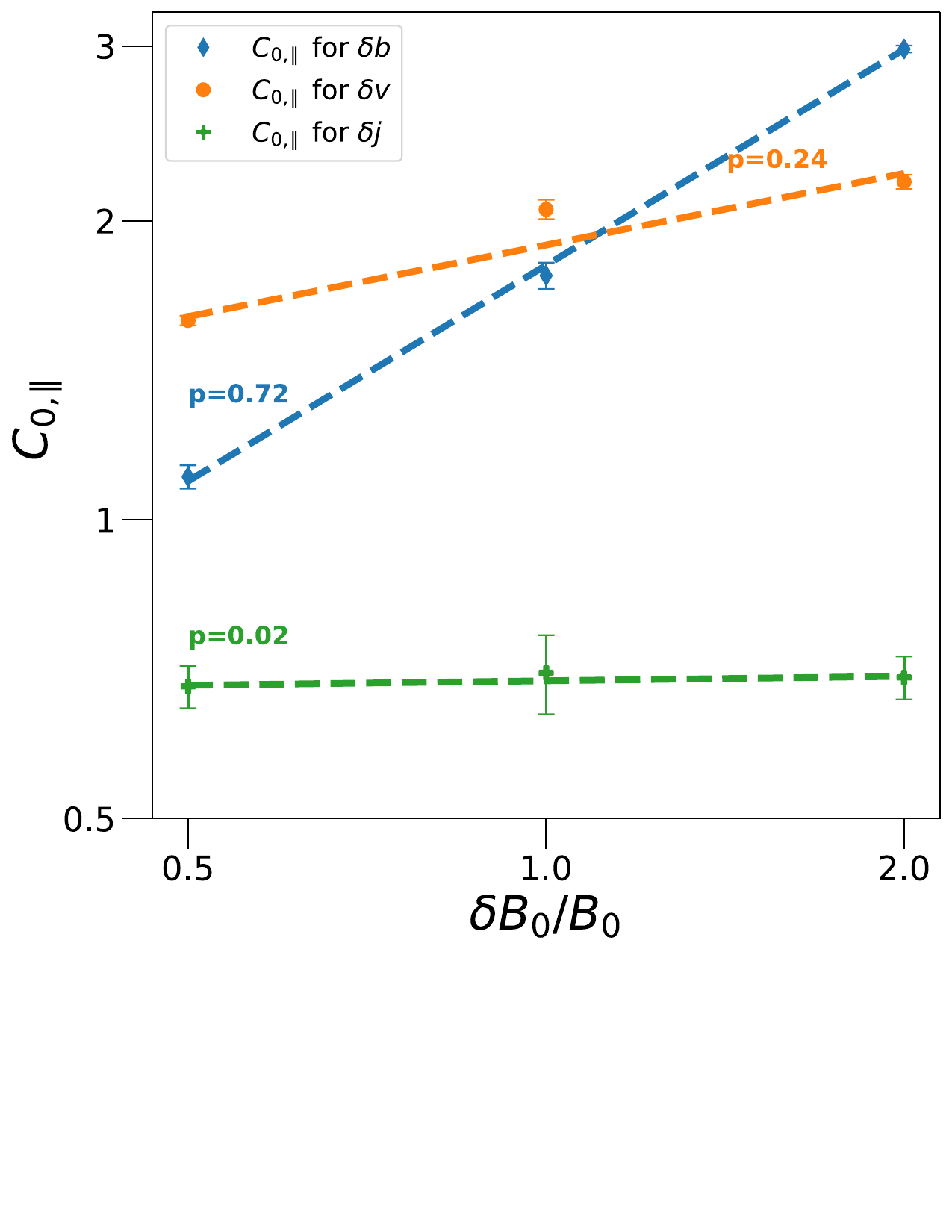} \\[\abovecaptionskip]
  \end{tabular}
   \vspace{-3.0cm}
  \caption{Dependence of the co-dimensions $C_0$ and $C_{0,\parallel}$ on ${\delta B_0}/B_0$. Orange, blue, and green denote fluctuations in $\delta v$, $\delta b$, and $\delta j$, respectively. Dashed lines show a power law with a similar slope to data of the corresponding color. Top and bottom figures are in log-log space}
  \label{fig:C0_vs_db}
\end{figure}
Figures \ref{fig:c0_vs_sigma} and \ref{fig:C0_vs_db} illustrate the behavior of the co-dimensions $C_0$ and $C_{0,\parallel}$ (for $\delta v$, $\delta b$, and $\delta j$) with respect to the magnetization $\sigma$ and the turbulence level ${\delta B_0}/{B_0}$. $C_0$ and $C_{0,\parallel}$ exhibit similar trends, albeit slightly higher values for the parallel component.
Figures \ref{fig:c0_vs_sigma} and \ref{fig:C0_vs_db} suggest an almost inverse trend in the fluctuations of velocity ($\delta v$) compared to magnetic field ($\delta b$), deviating from the expected scaling in incompressible MHD. This divergence might be attributed to the significant density fluctuations observed in relativistic turbulence. Future investigations could explore whether considering density-weighted fluctuations in velocity, as conducted in \citet{zrake2012}, could reinstate the incompressible MHD behavior.

As the dissipation of the velocity fluctuations go from filamentary to sheet-like with increasing $\sigma$, the magnetic fluctuations go from being dissipated in sheet-like structures to filamentary ones. This suggests that in increasingly magnetized plasmas, more energy is dissipated within flux rope-like structures, as seen in \citet{dong2022}. The general dependence on ${\delta B_0}/{B_0}$ is strong for the magnetic field and has a similar trend as is seen in \citet{muller2003}.

\section{Dissipative Structures}  \label{sec:diss_struct}

\begin{figure}
    \centering
    \includegraphics[width=\linewidth]{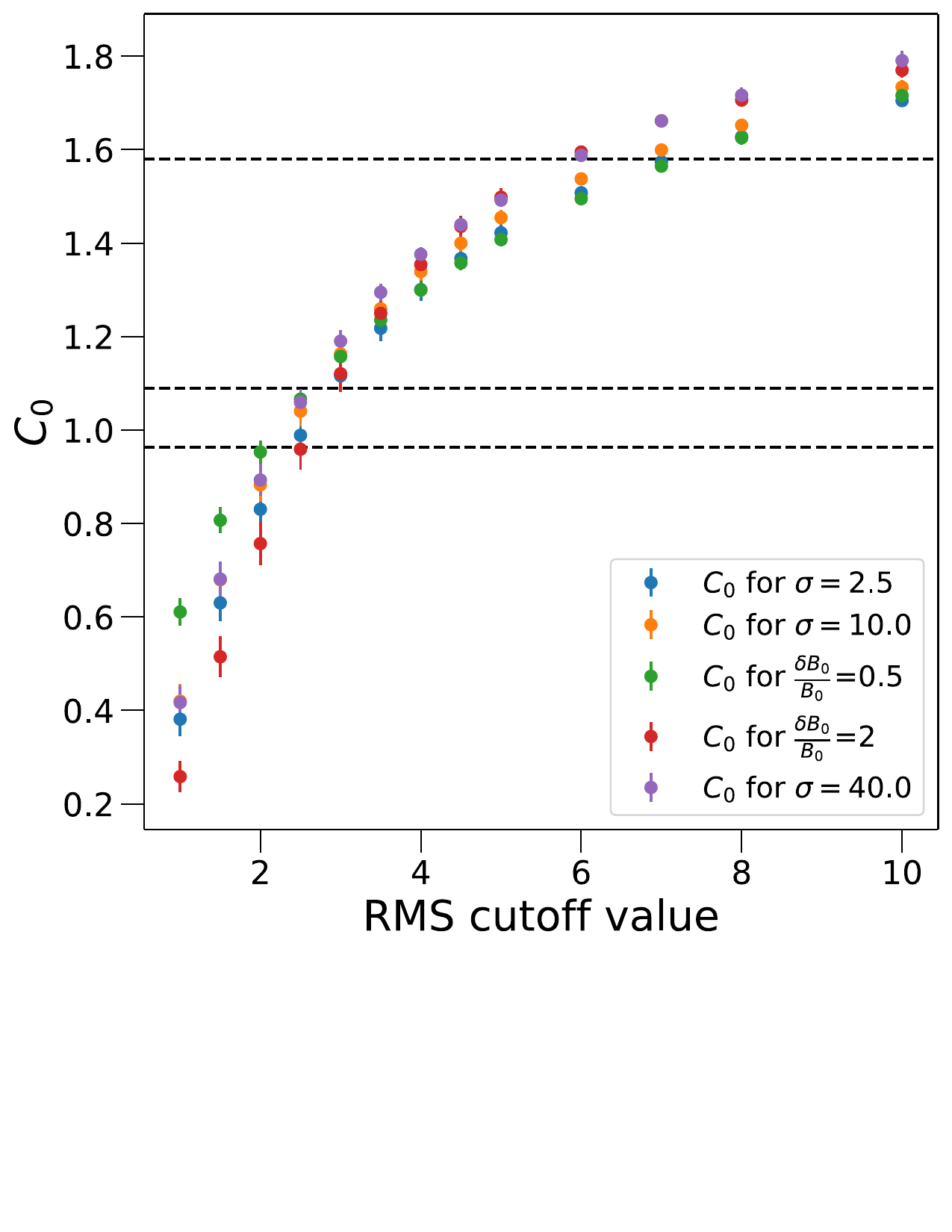}
    \vspace{-3.0cm}
    \caption{Co-dimension $C_0$ for $D_e \geq D_{e,{\rm rms}} \times RMS_{\rm cut}$ vs $RMS_{\rm cut}$. Horizontal dashed lines indicate $C_0$ values corresponding to $\sigma = 2.5, \, 10, \, 40$ as listed in Table 1. }
    \label{fig:De_vs_rmscutoff}
\end{figure}

In collisionless plasmas, a comprehensive theory explaining how the cascade eventually leads to dissipation in this regime is still lacking.
Given that energy dissipation must occur through electromagnetic interactions, and that energy is initially stored in the magnetic field, we choose to use the magnetic field cascade to define the co-dimension $C_0$ for the dissipative structures. We identify  structures with the corresponding $C_0$ by finding regions characterized by large electromagnetic dissipation. To this purpose, we employ the dissipation measure \citep{Zenitani2011} 
\begin{equation}
    \label{eq:D_e}
    D_e= \bm{J} \cdot\left(\bm{E}+\bm{v}_e \times \bm{B}\right) - \rho_e\left(\bm{v}_e \cdot \bm{E}\right) \, ,
\end{equation}
similarly to \citet{wan2016} in the context of nonrelativistic turbulence. Equation (\ref{eq:D_e}) describes the work done by electromagnetic fields on the particles, evaluated in a frame moving with the electron fluid velocity $\bm{v}_e $. Here, $\rho_e$ is the charge density and $\bm{E}$ is the electric field. In order to find the value of the co-dimension $C_0$ for structures in $D_e$, we first establish the fractal dimension for structures defined by $D_e/D_{e,{\rm rms}} \geq RMS_{\rm cut}$ using a box counting algorithm. We plot this for various values of $RMS_{\rm cut}$ in Figure \ref{fig:De_vs_rmscutoff}.
\begin{figure*}
  \centering
    \includegraphics[height=\textheight]{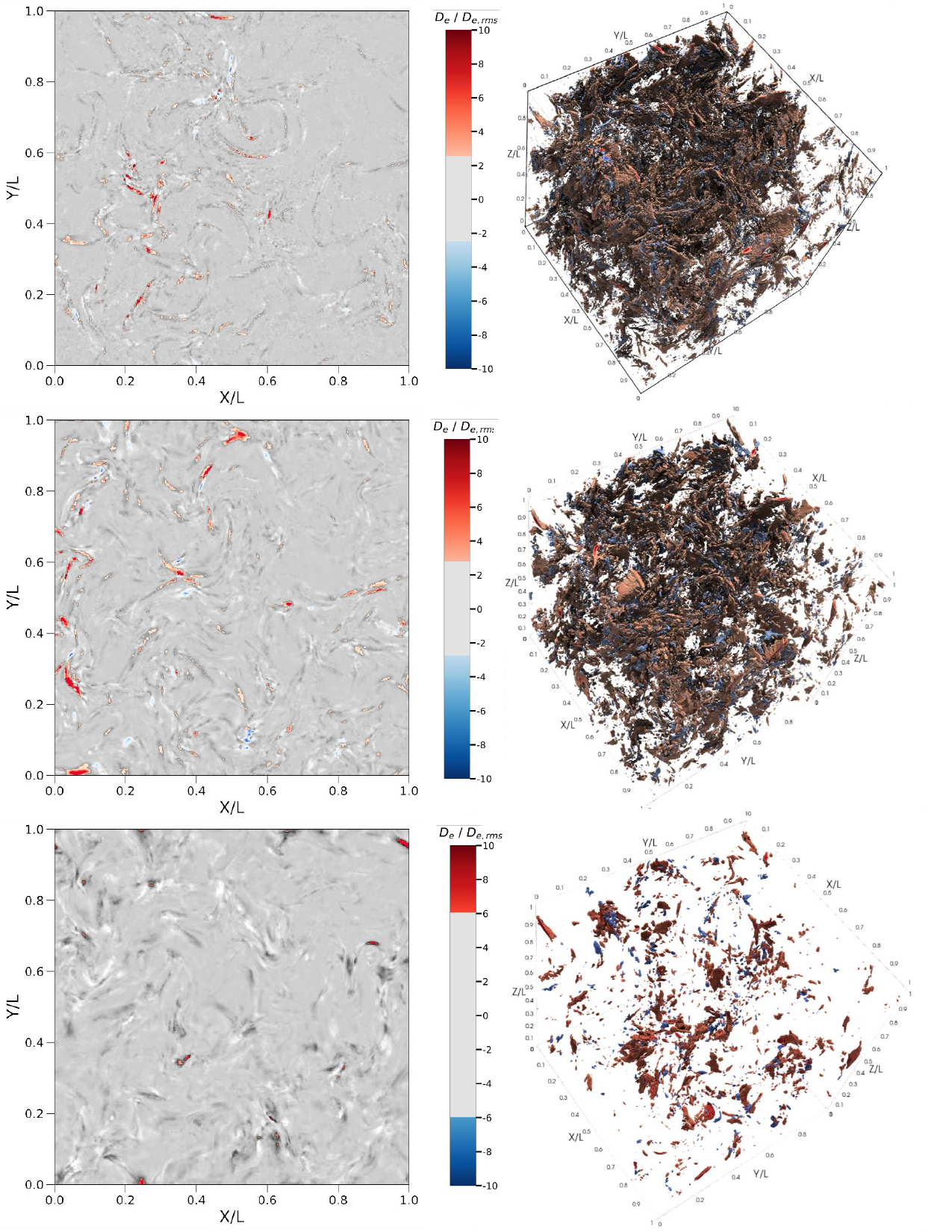}

  \caption{2D representation (left) and 3D representation (right) illustrating the most dissipative structures. From top to bottom, these structures correspond to turbulence with $\sigma = 2.5, \, 10, \, 40$. The figures show that as $\sigma$ increases, the coherent structures becomes increasingly dissipative while becoming less volume filling.}
  \label{fig:slices_and_3d}
\end{figure*}
From this plot we observe minimal dependence on $\sigma$ or  ${\delta B_0}/{B_0}$. Figure \ref{fig:De_vs_rmscutoff} enables us to visually identify the values of $RMS_{\rm cut}$ corresponding to a given co-dimension $C_0$, determined earlier by analyzing the SF. In Figure \ref{fig:slices_and_3d} we show a random 2D slice at a given $z$ coordinate. Structures are highlighted to show regions with a similar value of $C_0$ in $D_e$ as was found from fitting the SF exponent for $\delta b$. The 3D rendering of these structures is shown in the right column of Figure \ref{fig:slices_and_3d}. 

At low $\sigma=2.5$, we find a co-dimension $C_0 \approx 1$,  indicative of predominantly sheet-like structures in the dissipation measure $D_e$. These can be seen in Figure \ref{fig:slices_and_3d}(a) where discrete sheets of dissipation are discernible. In its corresponding 3D figure, we can see that these sheets fill a large part of the volume in comparison to other figures. At $\sigma=10$, the corresponding $C_0 \approx 1.1$ is highlighted in the $D_e$ structures with an $RMS_{\rm cut} \approx 3$. Though generally still sheet-like, the volume occupied by these structures is significantly reduced. For $\sigma=40$, the co-dimension $C_0$ increases to $\approx1.6$, but this requires a much higher $RMS_{\rm cut}\approx6$. This drastically decreases the occupied volume with small slightly elongated dissipation areas that resemble flux-ropes. It is clear that as $\sigma$ increase, the energy is dissipated in smaller, less volume-filling structures.

To quantify this result, for every simulation in Table 1, we used a bisection algorithm to determine the $RMS_{\rm cut}$ value that identifies structures defined by $\lvert D_e \rvert \geq \lvert D_{e,{\rm rms}} \rvert \times RMS_{\rm cut}$ with the same value for $C_0$ as listed in Table 1 within an error margin of $\pm 1\%$. Then, $f = N_{D_e}/N$ represents the filling fraction or the number of cells where $\lvert D_e \rvert \geq \lvert D_{e,{\rm rms}} \rvert \times RMS_{\rm cut}$.  Additionally, we define $q_f = {f_+}/{f_-}$, where $f_+$ is the filling fraction for structures defined by $D_e \geq D_{e,{\rm rms}} \times RMS_{\rm cut}$ and $f_-$ is the filling fraction for structures defined by $D_e \leq - \lvert D_{e,{\rm rms}} \rvert \times RMS_{\rm cut}$. The error bars for both $f$ and $q_f$ are derived from repeating these calculations using the upper and lower bounds of the original $C_0$ as reported in Table 1.
\begin{figure}
  \centering
  \begin{tabular}{@{}c@{}}
    \includegraphics[width=\linewidth]{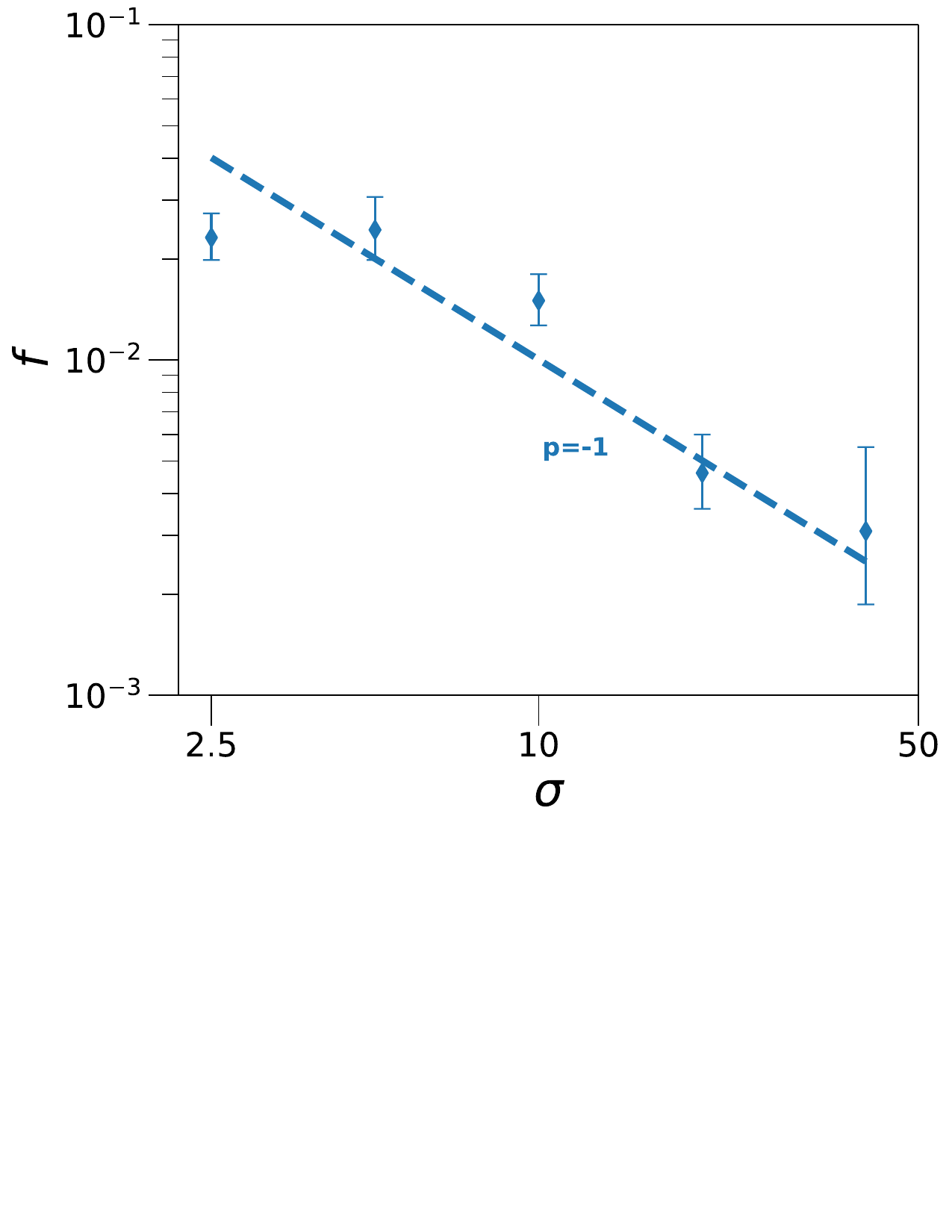} \\[\abovecaptionskip]
  \end{tabular}

  \vspace{-4.25cm}

  \begin{tabular}{@{}c@{}}
    \includegraphics[width=\linewidth]{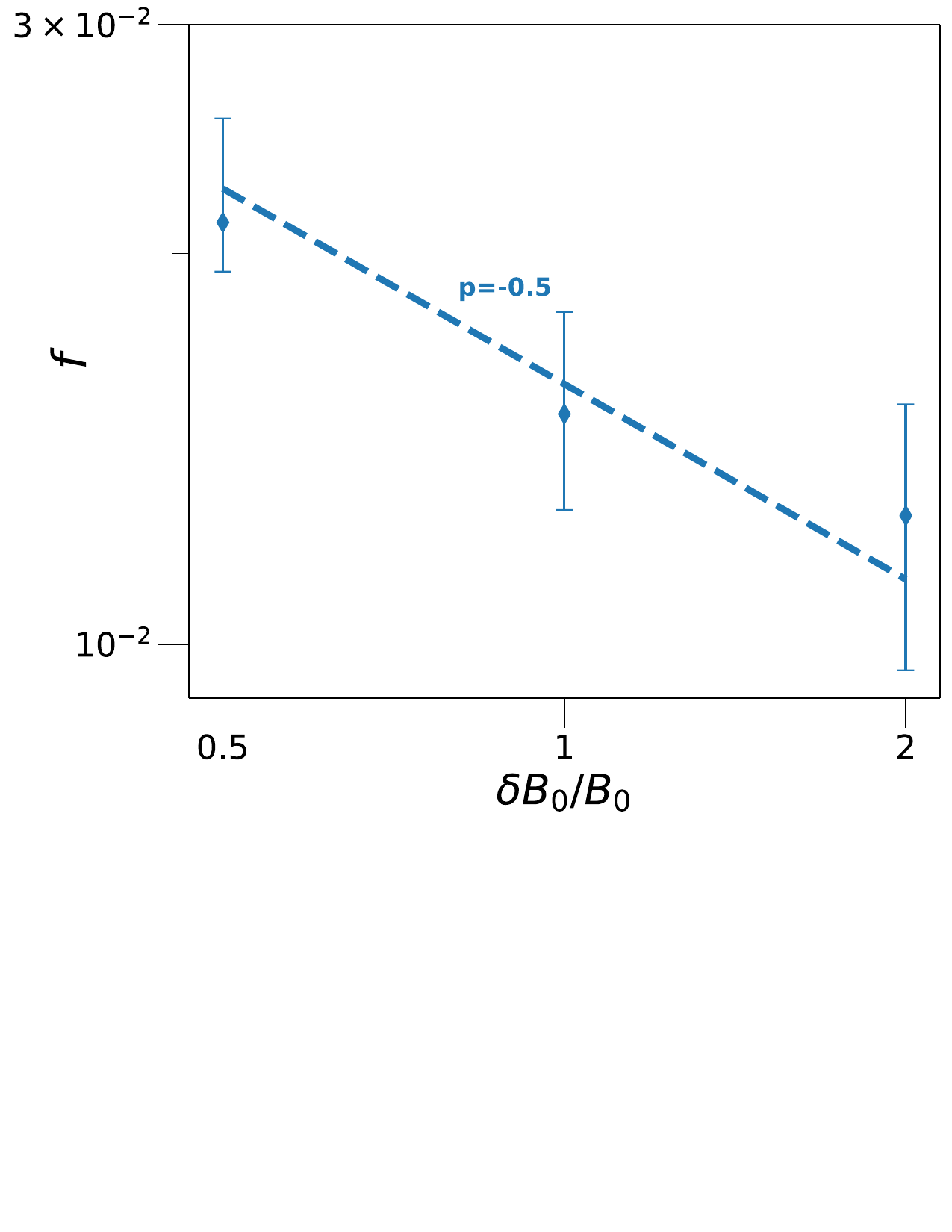} \\[\abovecaptionskip]
  \end{tabular}
  \vspace{-4.0cm}

  \caption{Top: Dependency of the filling fraction $f$ of the dissipative structures on the magnetization $\sigma$. Bottom: $f$ dependence on ${\delta B_0}/{B_0}$. Dashed lines show a power law with a similar slope to data.  }
  \label{fig:fill_vs_db}

\end{figure}
Results for $f$ can be seen in Figure \ref{fig:fill_vs_db} to be generally trending to smaller values before appearing to approach a constant as $\sigma$ increases. Larger values of ${\delta B_0}/{B_0}$ also result in decreased filling fraction of the inferred structures (Figure \ref{fig:fill_vs_db}).
\begin{figure}
  \centering
  \begin{tabular}{@{}c@{}}
    \includegraphics[width=\linewidth]{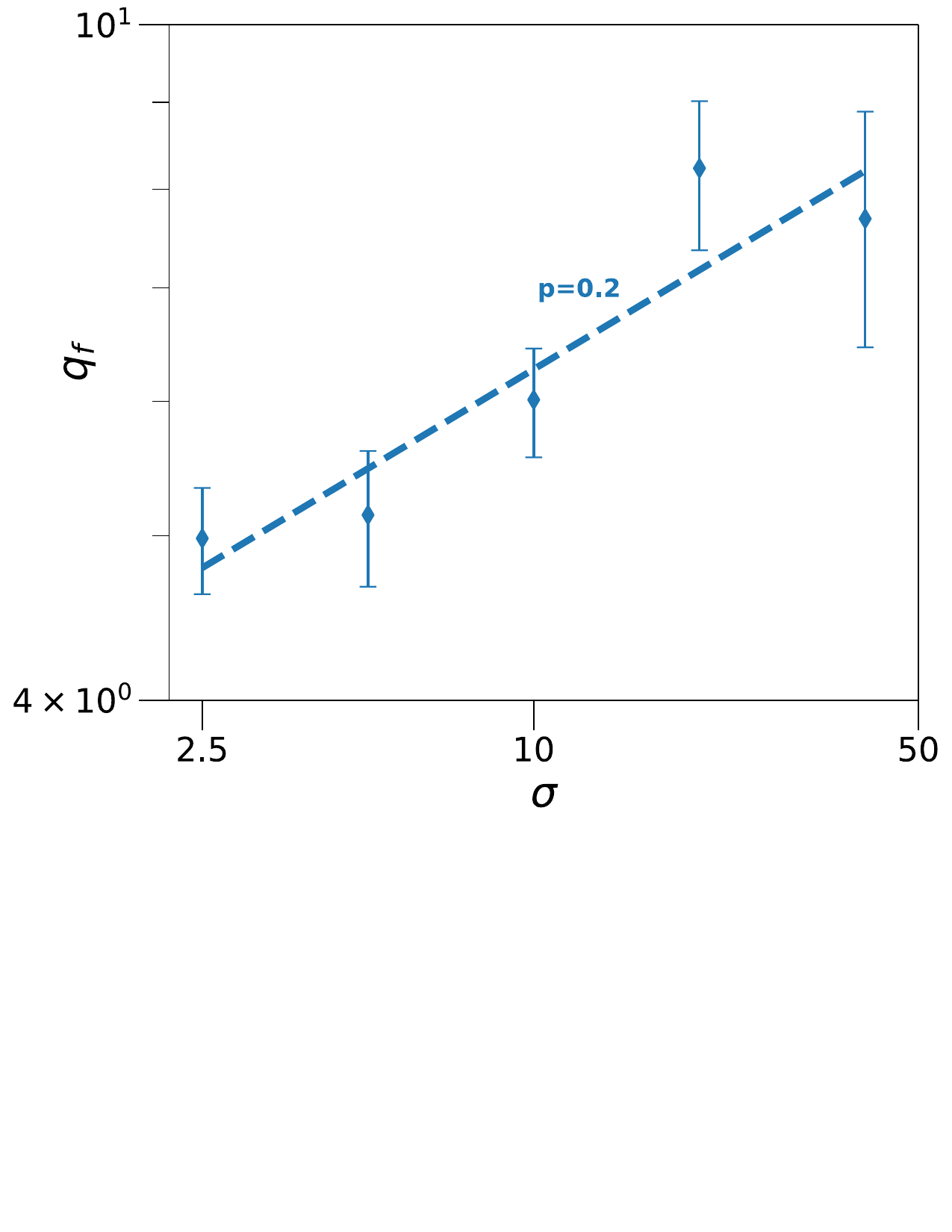} \\[\abovecaptionskip]
  \end{tabular}

  \vspace{-4.25cm}

  \begin{tabular}{@{}c@{}}
    \includegraphics[width=\linewidth]{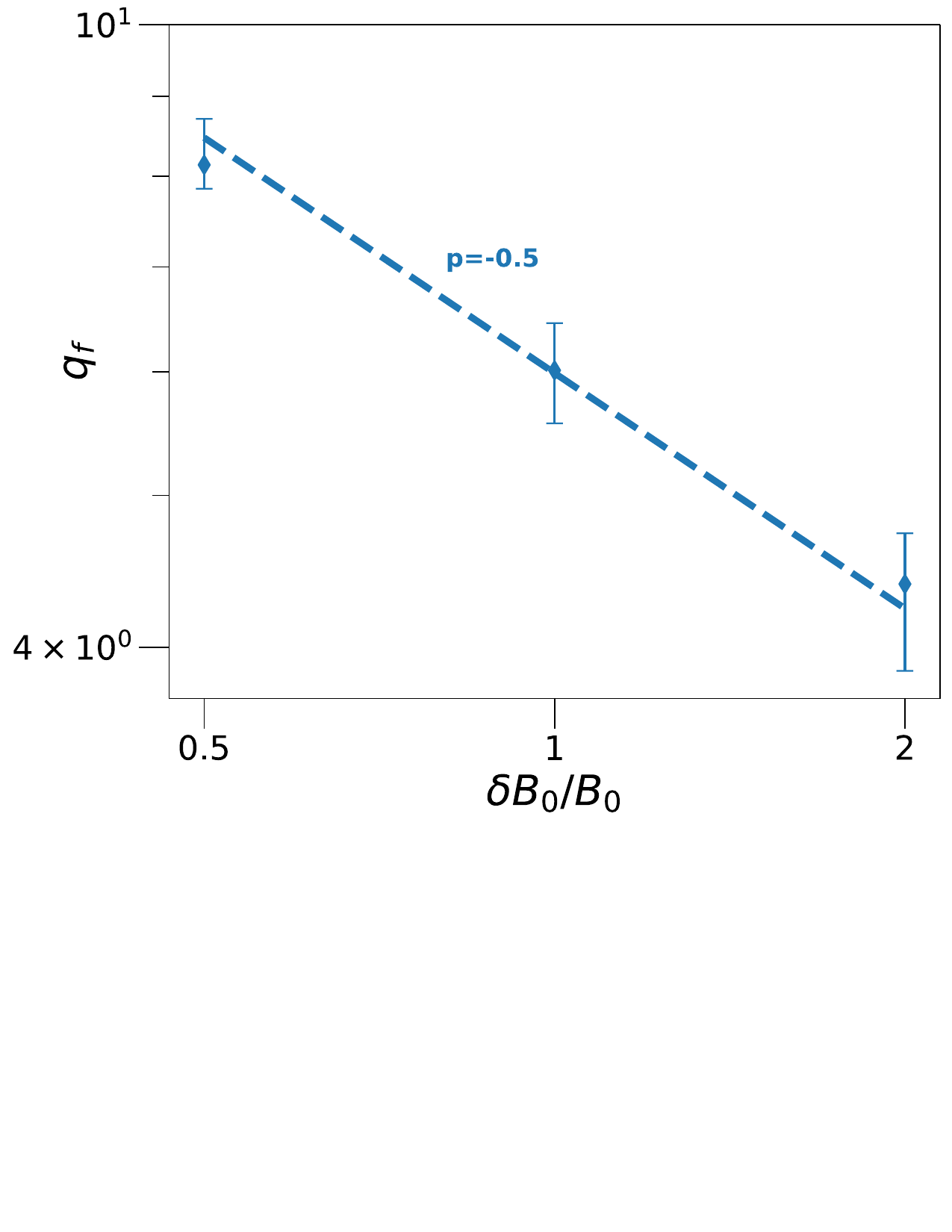} \\[\abovecaptionskip]
  \end{tabular}
  
\vspace{-4.0cm}
  \caption{Top: Dependence of $q_f = {f_+}/{f_-}$ on $\sigma$. Bottom: $q_f$ dependence on ${\delta B_0}/{B_0}$.  Dashed lines show a power law with a similar slope to data. }
  \label{fig:qfill_vs_db}
\end{figure}
The ratio $q_f$ appears in Figure \ref{fig:qfill_vs_db} to increase rapidly with $\sigma$ before it potentially plateaus. Additional data simulations will be required to determine if the trend continues for larger values of $\sigma$. The impact of ${\delta B_0}/{B_0}$ on $q_f$ is shown in the bottom frame of the same figure, revealing a steep decreasing trend.

\section{Discussion and Conclusions}  \label{sec:discussion}

When examining the identified dissipative structures, a trend emerges, showcasing a preference for more compact regions of dissipation (Figure \ref{fig:slices_and_3d}) as the magnetization $\sigma$ increases. This tendency may be attributed, in part, to current sheets becoming increasingly tearing unstable for larger values of $\sigma$ (and ${\delta B_0}/{B_0}$), resulting in the disruption of current sheets with a broader thickness $\lambda_d$. The disruption of these sheets enables the cascade to extend into smaller regions where the dissipation measure becomes stronger. In \citet{comisso2019}, $\lambda_d$ is derived and shown to have strong dependence on the plasma's enthalpy. Higher values of $\sigma$ increase the available free energy, leading to enhanced plasma heating and consequently an increase in $\lambda_d$. In future work, we aim to explore the potential link between the most dissipative structures and the formation of flux-ropes via tearing instability.

In \citet{comisso2018,comisso2019} it was shown that turbulent acceleration is a two stage process with the first being initial rapid acceleration associated with current sheets, and second being stochastic  Fermi-like process. The initial rapid acceleration works as an injection process for the particles and is controlled by the parallel electric field, thereby establishing a preferred direction of motion along the magnetic field for particles around this injection energy. If these same features investigated here predominantly operate through the parallel electric field, then understanding the distribution of these structures may provide a link to understanding the anisotropic features that arise in relativistic turbulence. Specifically, combining the distribution of coherent structures with the rate of plasma processed by a reconnecting current sheet, given by the reconnection rate  $\beta_R = v_R/c  \approx 0.1 \delta B_0/\sqrt{\delta B_0^2 + B_0^2} $ \citep{Comisso23}, we can approximate the average power per unit area experienced by the plasma from the parallel electric field as 
\begin{equation}
\label{eq:non-ideal_work}
P_\parallel = m c^2\int^{l_{\rm max}}_{l_{\rm min}} \frac{f}{1 + 1/q_f} \left(\frac{l}{l_{\rm min}} \right)^{C_0}    \gamma_{\rm inj} \ \beta_R \ dl \, ,
\end{equation}
where $l_{\rm min}$ refers to the minimum current sheet size, $ l_{\rm max}$ corresponds to largest scale current sheets, and $\gamma_{\rm inj} \propto \sigma$ \citep{comisso2019} is the mean particle Lorentz factor reached after being injected by the current sheet. 

PIC simulations have consistently demonstrated the two-stage nature of particle acceleration in magnetized turbulence \citep{comisso2018,comisso2019,Comisso21PRL,Nattila2021} and any successful theory of particle acceleration must encompass not only stochastic acceleration effects but also account for initial acceleration by non-ideal fields. \citet{comisso2018,comisso2019} showed via particle tracking that the highest energy particles usually first go through an injection phase that is associated with current sheets. If this is the case, then an acceleration theory for the injection phase requires both an understanding of how particles interacting with current sheets (or other dissipative structures) gain energy and how the distribution of dissipative regions depends on the initial plasma parameters. This work has focused on the latter, but a full acceleration theory will need to couple this with an adequate understanding of dissipation inside these structures. In \citet{lemoine2021}, the author reviews a random walk through intermittent structures as informed by large deviation theory. In the random walk with a time step $t_{\rm rw}$, a particle is assumed to have a probability $f_+$ of gaining a momentum fraction $g$, a probability $f_-$ of losing momentum fraction $g$ and a probability $1-f_+ - f_-$ of not interacting in that time step. This model showed it was capable of reproducing a hardened particle spectrum, but applying it here requires an understanding of how $g$ depends on $\sigma$ and $\delta B_0/B_0$. Further still, to use this model as an acceleration theory would require the assumption that the particle dynamics inside of magnetized turbulence can be reduced to a Levy flight.

In this work, we  combined intermittency analysis with phenomenological scaling theories to gain insights into the distribution of dissipative coherent structures within relativistic turbulent plasmas. Specifically, through the construction of  SFs up to order 10 from 3D relativistic turbulent PIC simulations, we accurately measured the SF exponents $\zeta_p$, which, when fitted to a generalized She-Leveque model, revealed the value of the co-dimension $C_0$ of the most dissipative structures.  We then determined the 3D representation of these structures by identifying regions of significant dissipation that shared the same value of $C_0$. Once identified, we were able to determine the filling fraction and the subset of positive to negative work structures. All of this was done for a range of $\sigma$ and ${\delta B_0}/{B_0}$ values, providing insights into how these properties scale within the regime of \textit{relativistic turbulence}. The value of the co-dimension $C_0$ for the magnetic field generally increases for both $\sigma $ and ${\delta B_0}/{B_0}$ going from 1, a value representative of a sheet, toward a value of 2, more filamentary like. Nonetheless, when pinpointing the most dissipative structures with larger values of $C_0$, they exhibit characteristics more akin to discrete flux-ropes rather than extended sheets. This suggests the possibility of current sheets becoming more tearing unstable, allowing the cascade to continue into smaller structures. Furthermore, we found that the filling fraction $f$ of the dissipative structures decreases with $\sigma$ and ${\delta B_0}/{B_0}$, plateauing at large values of $\sigma$. Additionally, the ratio of positive to negative work structures, $q_f$, increases with $\sigma$, leveling off at large $\sigma$ values, while decreasing with ${\delta B_0}/{B_0}$. These results contribute to understanding the link between scaling theories and intermittent coherent structures, paving the way for the development of a physically informed turbulent acceleration theory.

\begin{acknowledgments}
ZD and DG acknowledge support from the NSF AST-2107802, AST-2107806 and AST-2308090 grants. LC acknowledge support from the NASA ATP award 80NSSC22K0667. 
We acknowledge computing resources provided by the Innovative and Novel Computational Impact on Theory and Experiment (INCITE) program, using resources of the Argonne Leadership Computing Facility, which is a DOE Office of Science User Facility supported under contract DE-AC02-06CH11357.
\end{acknowledgments}

\bibliography{main}{}
\bibliographystyle{aasjournal}

\end{document}